# SURFACE PRESSURE AND SHEAR STRESS FIELDS WITHIN A FRICTIONAL CONTACT ON RUBBER

**Danh Toan Nguyen[1,2], Pierdomenico Paolino[1], M.C. Audry[1,3], Antoine Chateauminois[1], Christian Fretigny[1], Yohan Le Chenadec[2], M. Portigliatti[2], E. Barthel[3]**

[1] *Laboratoire de Physico-Chimie des Polymères et des Milieux Dispersés, UMR CNRS 7615, Ecole Supérieure de Physique et Chimie Industrielles (ESPCI), Université Pierre et Marie Curie (UPMC), Paris, France*

[2] *Manufacture Française des Pneumatiques Michelin, Centre de Technologies, Clermont-Ferrand, France*

[3] *Surface du Verre et Interfaces, UMR 125, CNRS / Saint Gobain, Aubervilliers, France*

**Abstract**

This paper addresses the issue of the determination of the frictional stress distribution from the inversion of the measured surface displacement field for sliding interfaces between a glass lens and a rubber (poly(dimethylsiloxane)) substrate. Experimental results show that high lateral strains are achieved at the periphery of the sliding contacts. As a consequence, an accurate inversion of the displacement field requires that finite strains and non linear response of the rubber substrate are taken into account. For that purpose, a Finite Element (FE) inversion procedure is implemented where the measured displacement field is applied as a boundary condition at the upper surface of a meshed body representing the rubber substrate. Normal pressure is also determined by the same way, if non-diverging values are assumed at the contact edge. This procedure is applied to linearly sliding contacts as well as on twisting contacts.

## Introduction

Due to practical applications as well as for fundamental issues, friction on elastomers has received a large attention over the years. However, the role played by adhesion, viscoelasticity and roughness in friction remains unclear. For the latter property, for example, it is generally assumed that the apparent contact area is not representative of the real contact area, but should be dependent on the normal load, the elasticity and the roughness parameters. As a consequence, rough surface would exhibit *local* interfacial friction which depends on the actual contact area, and thus on the *local* normal pressure. Though qualitative evidences for these theories have been obtained (see for example references [1], [2]), experimental data allowing a quantitative assessment of the models remains scarce. In most of the experimental friction studies, the friction force is measured for different normal loads, velocities, geometries… As total friction force is a property integrated over the whole contact area where the normal pressure is not homogeneous, evaluation of the different models is thus rather indirect.

In an effort to obtain a spatial resolution of the interfacial stress in the sliding contact of a rigid body on a flat rubber sample, we have recently developed a new technique, based on the measurement of the displacement field at the rubber surface [3], [4]. The distribution of the interfacial friction stress which causes the observed displacements is determined using a numerical inversion method. Field measurements methods are now very much used in the field of mechanics where they have emerged as a powerful technique to bridge the gap between experiments and simulations allowing for direct displacement and strain comparisons (for a review on that topic, see for example references [5] and [6]). One of the most popular approaches is the so-called Digital Image Correlation Technique (DIC) which can be viewed as an extension of the Particle Image Velocimetry (PIV) experiments widely used in fluid mechanics. In DIC measurements, the surface displacement field is determined from a comparison of the grey intensity changes of the object surface before and after deformation. As detailed in reference [6], various correlation algorithms can be implemented to determine the displacement field with progressive enrichment in mechanical information at the measurement stage. Field measurements are especially suitable in the case of heterogeneous mechanical measurements where spatial heterogeneities are not known a priori. In particular, they have been largely used in the field of fracture mechanics to investigate crack initiation and propagation. On the opposite, field measurements have only scarcely be applied to contact situations where spatial heterogeneities are also involved. One can mention a recent study by Scheibert *et al* [7], [8], where DIC approaches have been developed to obtain spatially resolved information about the frictional behaviour at an interface between a glass lens and a rough rubber surface during the incipient stages of sliding.

In this paper, we address the issue of the determination of the frictional stress distribution from the inversion of the surface displacement field for a contact between a glass lens and a flat rubber substrate. The effect of large strains on the accuracy of the inversion is especially considered. In the following, the experiments are briefly recalled. Two contact situations are considered: (*i*) a classical linear sliding configuration where a glass lens is rubbed on the PDMS substrate, (*ii*) a less conventional torsional contact configuration where the glass lens is twisted. In a previous investigation[3], we have established that torsional contacts are suitable for investigation of the failure of adhesion during stiction, i.e. incipient sliding stages. In the case of linear sliding, it is shown that, in order to obtain the surface shear stress field with a good accuracy, an hyperelastic model is to be used, as large deformations are involved in friction on rubber. The hyperelastic model is subsequently applied to the determination of the shear stress distribution during the stiction of twisting contacts.

**Materials and friction experiments**

A commercially available transparent poly(dimethylsiloxane) (PDMS) silicone (Sylgard 184, Dow Corning, Midland, MI) is used as an elastomer substrate. In order to monitor contact induced surface displacements, a square lattice of small cylindrical holes (diameter 8 µm, depth 11 µm) is produced on the PDMS surface by means of conventional micro-lithography techniques. Full details regarding the processing procedure are given in reference [3]. Under transmitted light observation conditions, this pattern appears as a lattice of dark spots. Their positions are easily detected using image processing. For the purpose of linear and torsional friction experiments, two kinds of surface lattices are generated which differ in the center to center spacing of the holes (40 µm and 400 µm for torsional and linear sliding experiments, respectively). Before use, the PDMS specimens are thoroughly washed with isopropanol and subsequently dried under vacuum. Millimeter sized contacts are achieved between the PDMS substrates and plano-convex BK7 glass lenses (Melles Griott, France) with a radius of curvature of ranging from 5.2 to 14.8 mm. The r.m.s. roughness of the lenses is less than 2 nm, as measured by AFM.

Friction experiments are carried out using two home-made devices described in references [3], [ 4]. Linear sliding experiments are performed under imposed normal load (between 1.3 N and 7.5 N) and velocity (between 0.01 and 1 mm s$^{-1}$). The PDMS substrate is moved with respect to the fixed glass lens by means of a linear translation stage. Under steady state sliding, images of the deformed contact zone are continuously recorded through the transparent PDMS substrate using a zoom and a CCD camera. This system is configured to a frame size of (1024x1024) pixels with frame rates ranging from 1 Hz to 10 Hz.

Contact torsion experiments are carried out under imposed normal displacement conditions (penetration depth δ=75±5 μm). During the experiments, the glass lens is rotated at imposed angular velocity (1 deg s$^{-1}$) using a motorized rotation stage. Before twisting the lens, a contact dwell time of 10 minutes is systematically observed in order to allow for the development of adhesion. During torsion, images of the deformed contact zone are continuously recorded through the transparent PDMS substrate using the same optical equipment as for linear sliding experiments.

**Results and discussion**

*Linear sliding*

Figure 1 shows an example of the surface displacement field under linear sliding condition. In-plane displacement components $u_x$ and $u_y$ were determined from a measurement of the distortion of the surface holes lattice. Here $y$ is the sliding axis and $x$ is perpendicular. As detailed in reference [4], image accumulation under steady state friction allows to generate a displacement field with a high signal to noise ratio and a considerably better spatial resolution (of the order of 7 μm) than the marker' spacing on the PDMS surface (400 μm). The distribution of the $u_y$ component reflects Poisson's effect: the PDMS surface is compressed along the sliding direction at the leading edge and stretched at the trailing edge. In-plane strain components can be derived from the measured displacement field. In order to account for potential finite strains a deformation gradient tensor, ***F***, is used which can be decomposed into the product of two second order tensors, ***R*** and ***V***:

$$\boldsymbol{F}=\boldsymbol{VR} \qquad (1)$$

This decomposition relies on the fact that any transformation of an element from the undeformed to the deformed configuration may be obtained by first rotating by ***R*** the element and applying a subsequent stretch, ***V***. The so-called left stretch tensor, ***V***, was calculated from the measured in plane surface displacement components. In figure 2, the profile of the corresponding logarithmic strain along the sliding direction and on the symmetry axis is plotted for two experiments carried out using different lens radii and normal loads. As expected, compressive and tensile strains are achieved at the front edge and the leading edge of the contact, respectively. The important feature is that quite high strain values are achieved at the periphery of the contact. Strain as high as 0.4 can be induced, which falls well beyond the small strain hypothesis. In addition, these strain levels lie well within the non linear range of the mechanical response of the used PDMS, as indicated by conventional tensile experiments (see inset in figure 2).

At this stage, one may ask to what extent these non linear strains affect the accuracy of the inversion of the displacement field using a linear elastic approach such as the Green's tensor based procedure developed in reference [4]. At first sight, one may argue that non linearities are confined to the periphery of the contact and that they should have only a limited impact on the determination of the shear stress distribution inside the contact zone. This issue was addressed by means of finite element (FE) simulations which are able to handle large strain calculations with neo-Hookean materials (details on the FE calculations are provided in the appendix). Using the FE method, a displacement field was simulated which corresponds to the application of a constant shear stress, $\tau_0$, on a circular disk (in the undeformed configuration) at the surface of a neo-Hookean body. Then, the calculated lateral surface displacement field was inverted using the Green's tensor, i.e. under the assumption of small strain behaviour. When doing so, it emerges that the calculated shear stress distribution differs significantly from the constant shear distribution used to generate the displacement field (Figure 3). A clear dissymmetry is observed: shear stresses at the trailing edge are overestimated and, conversely, those at the leading edge are underestimated. Although localized at the periphery of the contact, large strains thus affect the whole contact response and the determination of shear stress inside the contact zone. One may consider that high strains at the edge of the contact result in the formation of an annular zone where the stiffness of the PDMS material is locally modified. As a result, the mechanical coupling between the external and inner parts of the contact is modified and the effects of large strain thus extend inside the contact zone.

These results of the FE analysis point out the need for a finite strain contact model in order to achieve a good accuracy in the determination of the frictional stress distribution from measured displacements. For that purpose, a FE inversion procedure of the displacement field was developed. The approach consists in prescribing the measured displacement field at the surface of a meshed FE body simulating the PDMS specimen and to retrieve the corresponding surface stress distribution under the assumption of large strain, neo-Hookean behaviour. Then, the local contact pressure and the frictional shear stress are calculated at every location within the contact zone from a projection of the components of the calculated surface stress tensor in a local Cartesian coordinate system. Its orientation is defined from the normal to the lens surface and from the actual sliding direction. As a consequence, the inversion procedure takes into account the contact geometry together with the measured sliding paths trajectories. In its spirit, this projection is similar to the flow line model developed by Lafaye *et al* [9] in the context of scratch experiments on polymer surfaces.

In a small strain situation, the inversion can be accurately carried out using only the in plane displacement components as it can be shown that the displacements induced by the normal and lateral loading components are fully decoupled in the case of an incompressible material [10]. This decoupling does not hold in the large strain regime and the FE inversion has therefore to be carried out using both vertical ($u_z$) and lateral ($u_x$ and $u_y$) surface displacements. Vertical displacements are not measured during sliding experiments, but they are prescribed within the contact zone by the known curvature of the lens and the unknown penetration depth, $\delta$. Interestingly, the later can be determined if some constraints are introduced regarding the normal stress distribution. Following some arguments by Savkoor [11], [12], it can be assumed that sliding prevents adhesion to develop quasi-singular strains near the contact edge as it is the case for static contacts. Accordingly, one may expect that normal stress should vanish at the periphery of the frictional contact. Then, this hypothesis allows the penetration depth, $\delta$, to be determined using an iterative procedure where the FE inversion of the displacement is carried out for a set of $\delta$ values until the calculated normal stress vanishes at the periphery of the contact. Figure 4 shows an example of the normal stress distributions along the *y* axis calculated for various indentation depths with the same measured lateral displacement field. The normal stress is seen to vanish for an indentation depth which is close to the theoretical Hertzian value, $\delta=a^2/R$, where *a* is a typical value of the contact radius (the contact region is not perfectly circular) and *R* is the curvature radius of the slider. When the penetration depth is either decreased or increased from this value, a stress peak (either positive or negative) is calculated at the periphery of the contact. The later can be assimilated to the stress singularity induced by a flat punch displacement component. In that sense, it is equivalent to the flat punch term introduced in the JKR analysis to account for adhesion [13]. Interestingly, an Hertzian value for the penetration depth was assumed by Savkoor in his description of the sliding contact.

The extent of coupling between normal and lateral contact components can be evaluated from a comparison of a set of two different inversions. The first one is carried out by prescribing the measured lateral displacement and $u_z=0$. In the second one, the three displacement components are prescribed at the surface of the FE specimen. The results show (figure 5) that a distinct, although limited, coupling exists between the normal and lateral contact loading components.

The large strain inversion procedure was subsequently applied to the investigation of the normal load and sliding velocity dependence of the local shear stress at the interface between PDMS and the smooth glass lens. Within the investigated contact pressure and velocity range, shear stress distribution was systematically characterized by a dissymmetry between the front and the rear part of the contact: as shown in figure 6, shear stress in the rear part of the contact is slightly increased as compared to that in the front part. On the opposite, a much more limited dissymmetry is noted on the normal stress profile. This dissymmetry in the shear stress profile is enhanced when the sliding velocity is decreased (figure 7). This effect could tentatively be accounted for by some viscoelasticity within the PDMS substrate, even if linear viscoelastic measurements show almost perfect elastic response at room temperature for a loading frequency **F**≈ $v/2a$, where $v$ is the sliding velocity and $a$ is the contact radius (tan $\delta$ = 0.04 at 25°C and 0.17 Hz). When the contact load is increased from 1.3 to 7.5 N (i.e. the mean contact pressure from 0.25 to 0.42 MPa), the shear stress distribution remains nearly unchanged (figure 8) which confirms previously reported results for a similar glass/PDMS contact [3], [4]. As a conclusion, the inversion of friction induced displacement field for a smooth glass/PDMS interface allows to identify a local friction law which is essentially independent on the contact pressure.

*Torsional contacts*

Surface displacement fields obtained under torsional contact conditions were detailed in a previous investigation. During the incipient stages of the frictional process, the stiction of the adhesive contact is associated with the progressive propagation of a slip annulus from the periphery of the contact This feature is shown in (Figure 9(I)) where the orthoradial displacements, $u_\theta$, is plotted as a function of the radial coordinate for various the twist angles. During stiction (i.e. for twist angle less than less than about 0.35 rad), the partition of the contact zone in an inner adhesive zone (where $u_\theta$ depends linearly on the radial coordinate, $r$) and an external micro-slip zone is clearly seen. Similarly to the linear sliding case, elevated shear strains ($du_\theta/dr|_{r=a}$≈ 0.4) are achieved at the vicinity of the contact edge which also question the accuracy of an inversion using a linear elastic analysis. Figure 9(II) shows the shear stress distribution obtained from the inversion of measured orthoradial displacement fields at various stages of the stiction process (radial displacements were found to be negligible). The inversion was carried out using FE hyperelastic calculations. In the friction zone, a nearly constant, pressure independent shear stress is achieved which is consistent with frictional behaviour under linear sliding. At the boundary between the stick and the slip zone, a stress overshoot is also retrieved which is indicative of the crack like nature of adhesive failure. These conclusions were already drawn using Green's tensor approach. This geometry appears particularly suitable for mode III crack propagation analysis. More detailed quantitative analysis of the experimental results is in progress.

**Conclusions**

A method for the determination of the local surface stresses in a sliding contact on rubber is presented. From the measurement of the surface displacement field, it is shown that FE calculation leads to a quantitative evaluation of both frictional shear stress and normal pressure. The high values of lateral strains which are measured for glass/rubber friction make it necessary to use of hyperelastic, finite deformation, calculations. This technique opens the way for a complete analysis of the local friction laws of rough contacts, which remains largely unexplored experimentally. It can also provide some insights into stiction processes of adhesive contacts. The usual route in the theoretical analysis of such a problem is to use a fracture mechanics approach where the boundary of the stick zone is assimilated to a crack. The present analysis could help in identifying any interplay between normal pressure and mode II / mode III crack propagation at the periphery of this moving boundary between stick and slip zones.

**Acknowledgements**

Part of this work was supported by the National Research Agency (ANR) within the framework of the DYNALO project (project N° NT09_499845). We also acknowledge financial support from Saint Gobain Recherche, Aubervilliers, France. We are also indebted to A. Prevost, G. Debrégeas (Ecole Normale Supérieure, Paris) for stimulating discussions about this work.

**References**

1. Persson, B.N.J., *J. of Chem. Phys.* **115**, 3840-3861 (2001).
2. Persson, B. N. J., Albohr, O., Tartaglino, U., Volokitin, A. I. and Tosatti, E., *J. Phys. Cond. Matter* **17**, R1-R62 (2005).
3. Chateauminois, A., Fretigny, C. and Olanier, L., *Phys. Rev. E* **81**, 026106 (2010).
4. Chateauminois, A. and Fretigny, C., *Eur. Phys. J. E* **27**, 221-227 (2008).
5. Pan, B., Qian, K., Xie, H. and Asundi, A., *Meas. Sci. and Tech.* **20**, 062001 (2009).
6. Hild, F. and Roux, S., *Strain* **42**, 69-80 (2006).
7. Scheibert, J., *Mécanique du contact aux échelles mésoscopiques.*, PhD Thesis, 2008, Ecole Normale Supérieure, Paris.
8. Scheibert, J., Debrégeas, G., and Prevost, A., *arXiv:0809.3188v1 [cond-mat.soft]*, 2008.
9. Lafaye, S., Gauthier, C. and Schirrer, R., *Trib Int* **38**, 113-127 (2005).
10. Landau, L.D. and Lifshitz, E.M., *Theory of elasticity,* (Pergamon, Oxford, 1986). 3$^{rd}$ ed., Vol. 7, pp.22-26.
11. Savkoor, A.R., *Models of friction based on contact and fracture mechanics*, in *Fundamentals of Friction: Macroscopic and Microscopic Processes*, Singer, I. and Pollock, H. Eds, (Kluwer Academic Publishers, Dordrecht, 1992). p. 111-133.
12. Savkoor, A.R. and Briggs, G.A., *Proc. Roy. Soc. London A* **356**, 103-114 (1977).
13. Johnson, K., Kendall, K. and Roberts A.D., *Proc. of the Roy. Soc. London A* **324,** 301-313 (1971).

**Appendix**

A commercially available finite element code (Abaqus 6.9-EF1) is used for all the numerical contact simulations and the inversion of experimental displacement fields. The PDMS substrate was modeled as a parallelepiped with the size of the sample 60x30x15 mm$^3$. Large strain simulations are carried out assuming a hyperelastic behaviour. Within the experimental strain range, tensile test showed that the behaviour of the PDMS material can be adequately described using a neo-Hookean's law. Accordingly, the later is used in the FEM simulations with the following two parameters: $C10=0.45$ MPa (i.e. half the shear modulus) and $D1=0.001$ MPa$^{-1}$ (i.e. twice the compliance). In order to account for the quasi-incompressibility of the material in the numerical analysis, linear hybrid quadrangular elements (C3D8RH) are used. The mesh was refined in the contact region where the surface element' size is 0.1x0.1x0.007 mm$^3$. The dimensions of the elements were selected from a preliminary study of mesh's convergence. In the inversion procedure, experimental displacements are applied to surface' nodes within a restricted region with the size of the measurement field. Normal displacements are applied only to the nodes inside the contact zone. All the other nodes are free except those located on the bottom face of the PMDS model where the displacements are set to zero. The calculations are performed using a non linear static analysis.

**Captions to figures**

**Figure 1**: Measured surface displacement field for linear sliding. (*a*) displacement along the sliding direction $u_y$, (*b*) displacement perpendicular to the sliding direction $u_x$. Radius of the contact lens, $R$=9.3 mm, normal load, $P$=1.52 N, sliding velocity $v$=0.7 mm s$^{-1}$. The PDMS substrate is moved from bottom to top with respect to the fixed glass lens as indicated by the arrow.

**Figure 2**: Profile of the logarithmic surface strain along the sliding direction and in the contact midplane for two different contact conditions. Thick line: $R$= 5.2 mm, $P$=1.67 N; plain line: $R$=9.3 mm, $P$=1.52 N ($v$=0.5 mm s$^{-1}$). Inset: tensile behaviour of the PDMS rubber (crosshead speed: 0.08 mm s$^{-1}$).

**Figure 3**: Normalized shear stress profiles obtained from the small strain inversion of a displacement field calculated by FEM in the case of a constant shear stress, $\tau_0$, applied over a circular region at the surface of a rubber substrate (see [4] for details). FEM simulations were carried out under the assumption of (*i*) linear elasticity, case (a) and (*ii*) large strains and neo-Hookean behavior of the rubber, cases (b) $\tau_0$= 0.1 MPa, (c) $\tau_0$= 0.2 MPa and (d) $\tau_0$= 0.3 MPa. Shear stresses are normalized with respect to $\tau_0$. The dotted line corresponds to the prescribed shear stress profile. Deviations from this distribution point to the need to take into account large deformations and non linearity in the deconvolution procedure.

**Figure 4** Contact pressure profiles along the *y* axis obtained from the FEM inversion of a surface displacement field using different assumptions regarding the value of the normal penetration depth, $\delta$. Dotted line: $\delta$=0.5 $a^2/R$; plain line: $\delta$= $a^2/R$; thick line: $\delta$= 1.5 $a^2/R$, where $R$ is the lens radius and $a$ is a typical value of the contact radius *($P$=1.52 N, $v$=0.7 mm s$^{-1}$)*.

**Figure 5** Shear stress profiles along the sliding direction as obtained from FEM inversion of an experimental surface displacement field. Bold line: profile obtained using only the measured $u_x$ and $u_y$ components ($u_z$=0). Plain line: profile obtained by specifying the three displacement components $u_x$, $u_y$ and $u_z$ ($P$=1.52 N, $R$= 9.3 mm, $v$=0.7 mm s$^{-1}$).

**Figure 6** FE inversion of the surface displacement field within a sliding contact between a smooth glass sphere and a PDMS substrate ($R$=9.3 mm, $P$=1.52 N, $v$=0.7 mm s$^{-1}$). (a) Distribution of the contact pressure and shear traction; (b) shear stress (bold line) and normal pressure (normal line) profiles across the contact zone and along the sliding direction.

**Figure 7** Surface shear stress profiles along the sliding direction for various sliding velocities. Dotted line: $v$=0.02 mm s$^{-1}$, plain line $v$= 0.1 mm s$^{-1}$, bold line: $v$=1 mm s$^{-1}$ ($P$=1.52 N, $R$=9.3 mm).

**Figure 8** Contact pressure (a) and surface shear stress (b) profiles along the sliding direction for $P$=1.3 N (plain line) and $P$= 7.5 N (bold line). $v$=0.5 mm s$^{-1}$, $R$=9.33 mm.

**Figure 9** Inversion of the surface displacement field induced by the torsion of a glass lens. (I) Orthoradial displacement measured for various angles of twist, $\theta$, during stiction. (II) Surface shear stress distribution obtained from the FEM inversion of the orthoradial displacement profiles. (a) $\theta$=0.029 rad; (b) $\theta$=0.087 rad; (c) $\theta$=0.175 rad, (d) $\theta$=0.262 rad, (e) $\theta$=0.349 rad (angular displacement rate: 1 deg s$^{-1}$, $R$=14.8 mm).

**Figure 1**

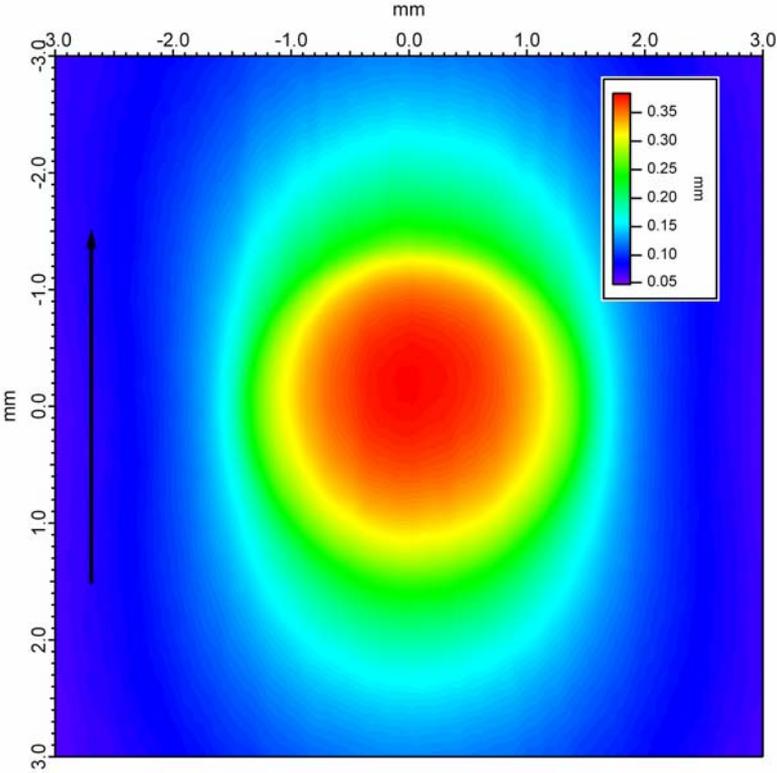

(a)

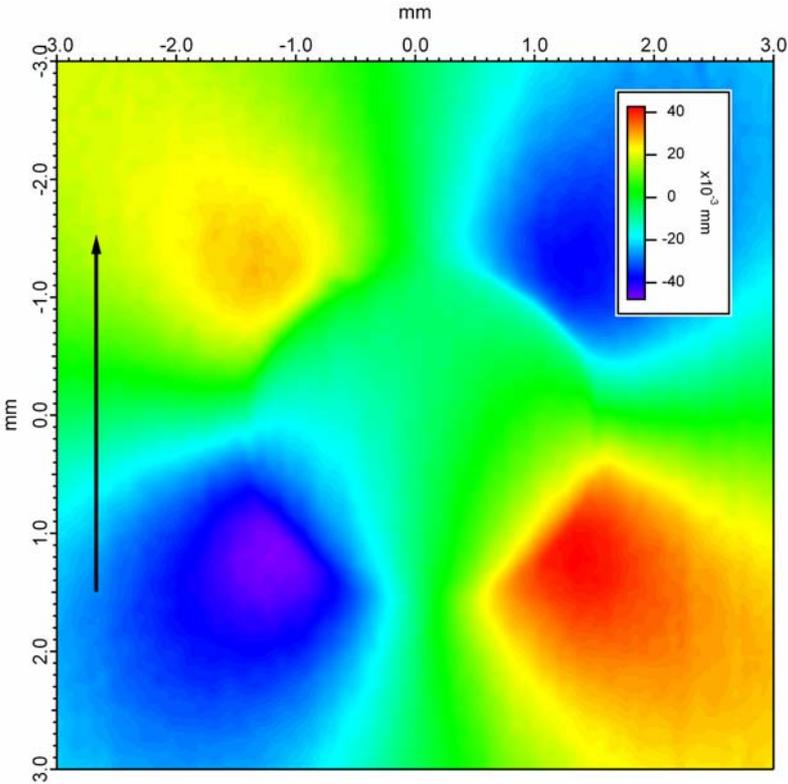

(b)

**Figure 2**

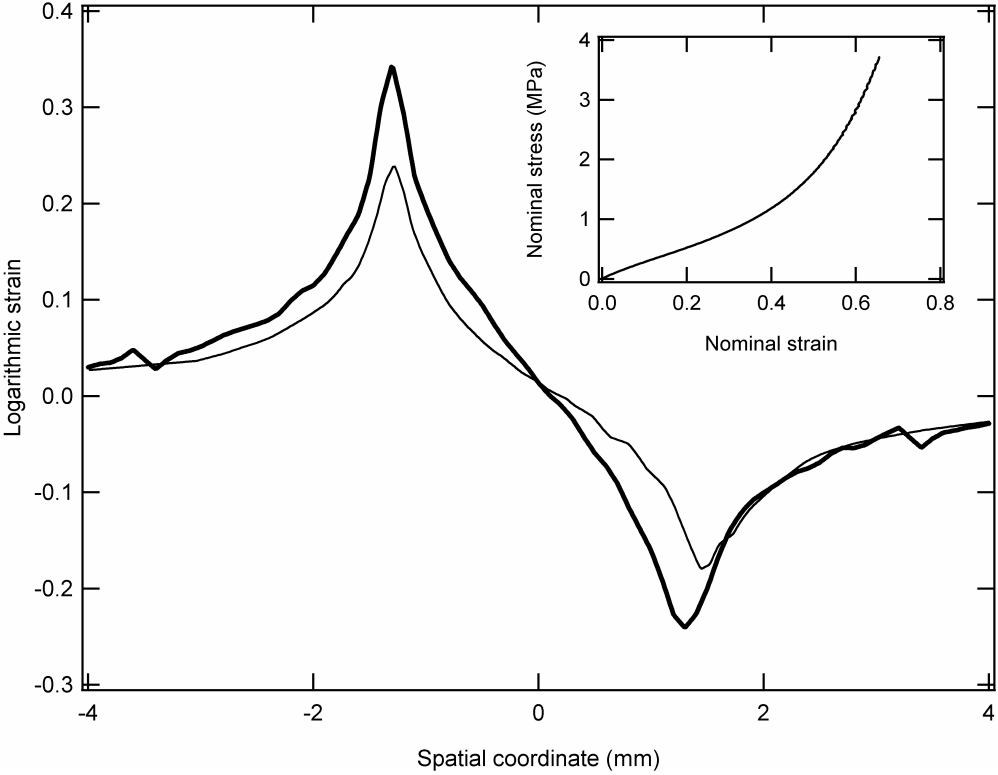

**Figure 3**

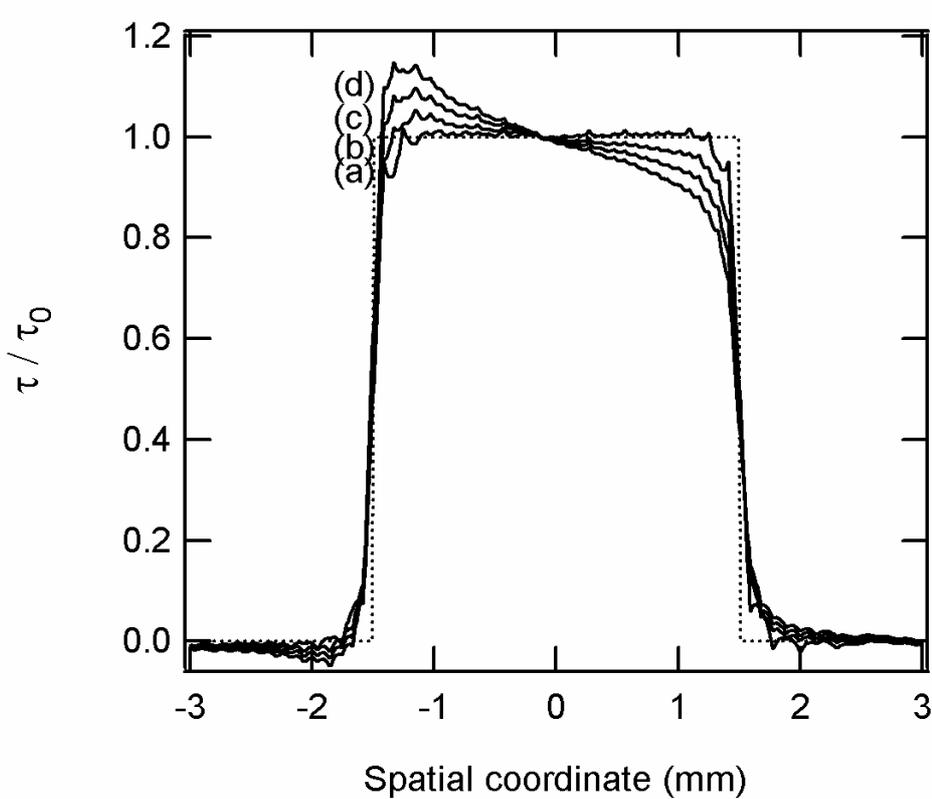

**Figure 4**

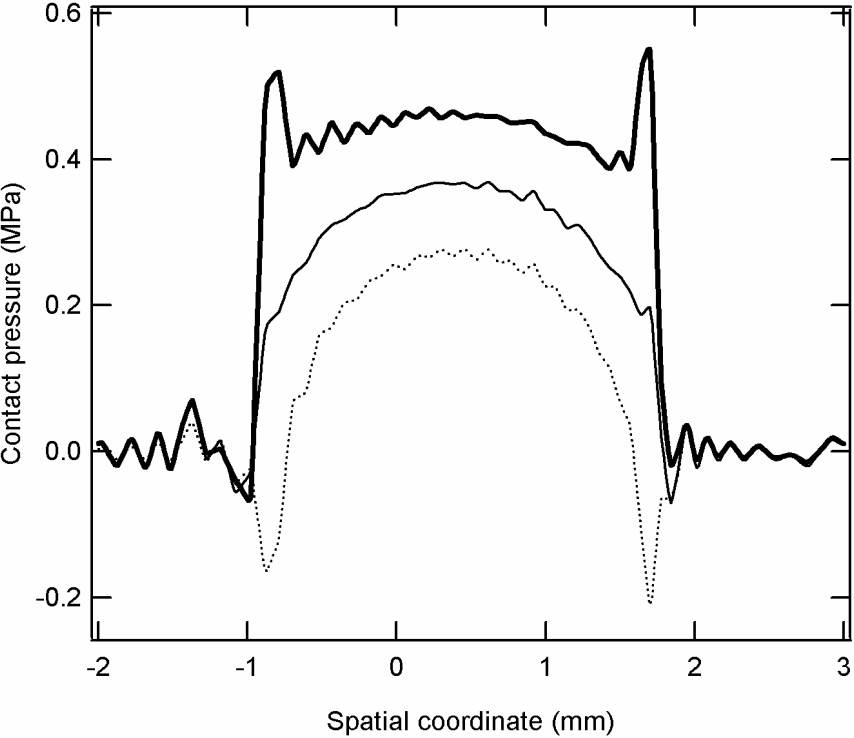

**Figure 5**

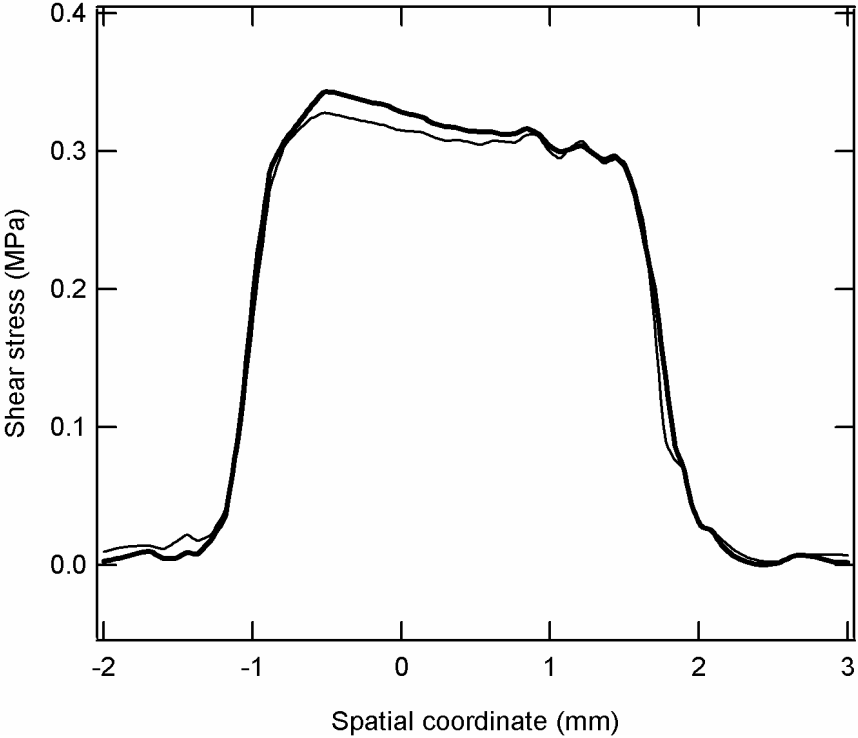

**Figure 6**

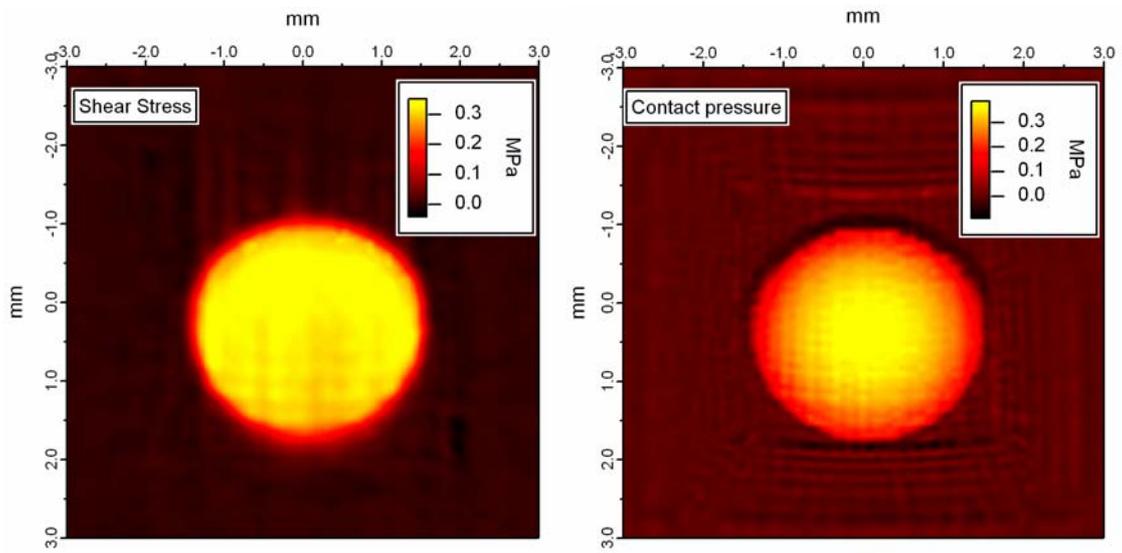

(a)

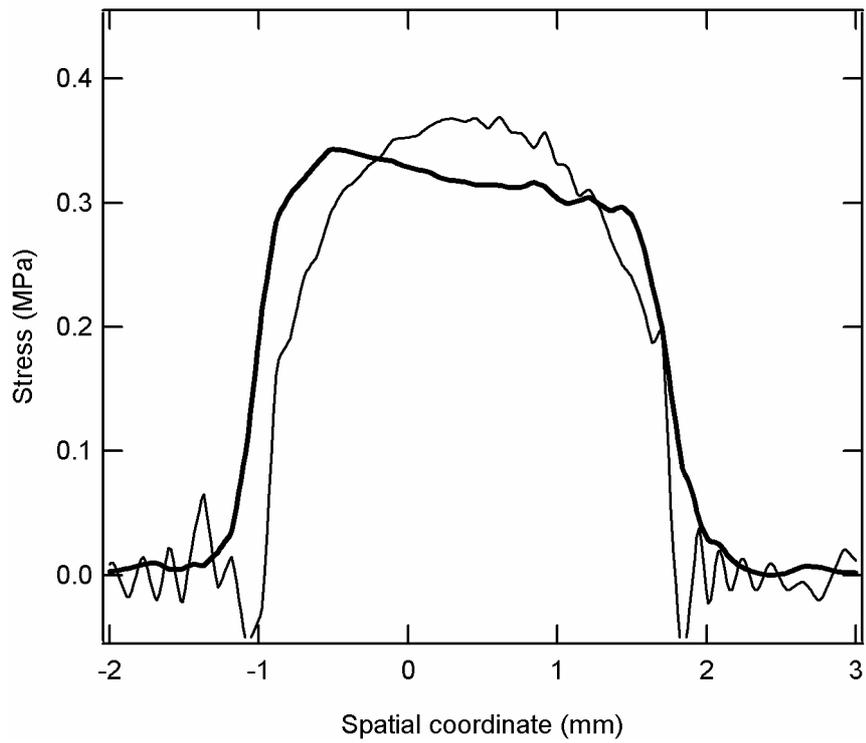

(b)

**Figure 7**

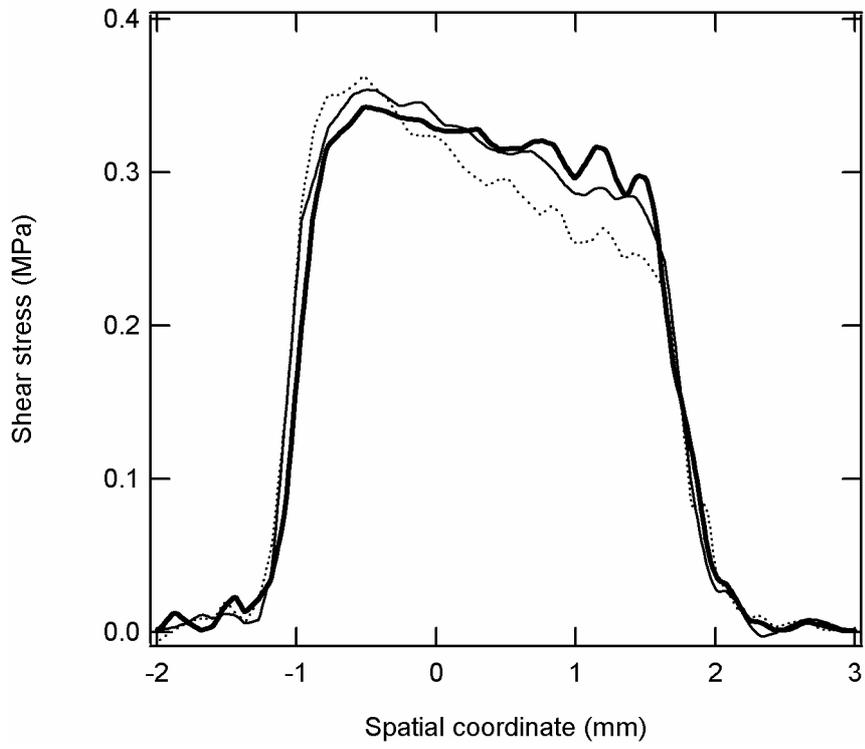

**Figure 8**

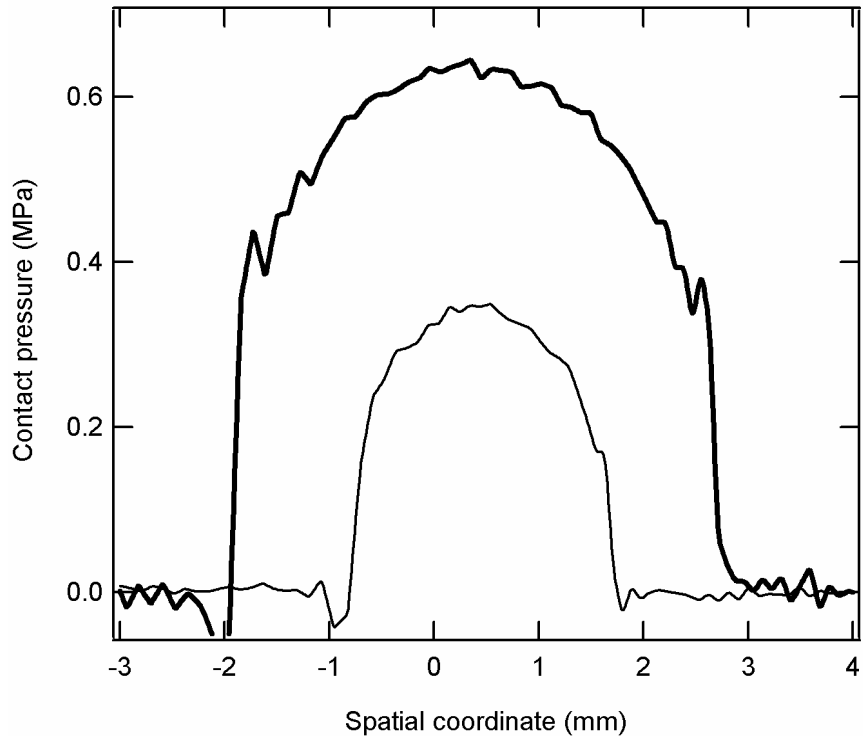

(a)

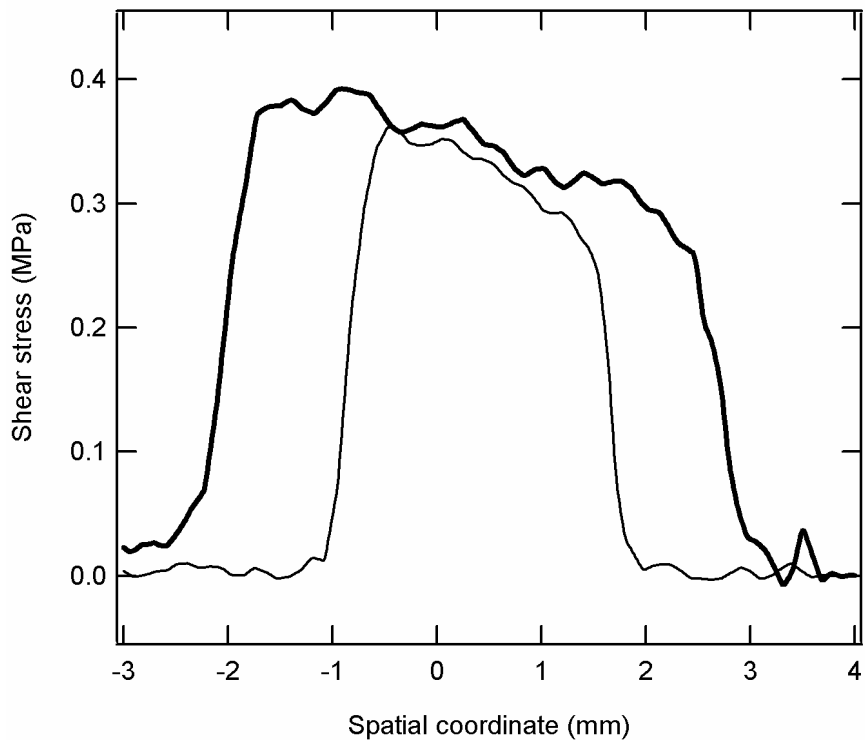

(b)

(b)

**Figure 9**

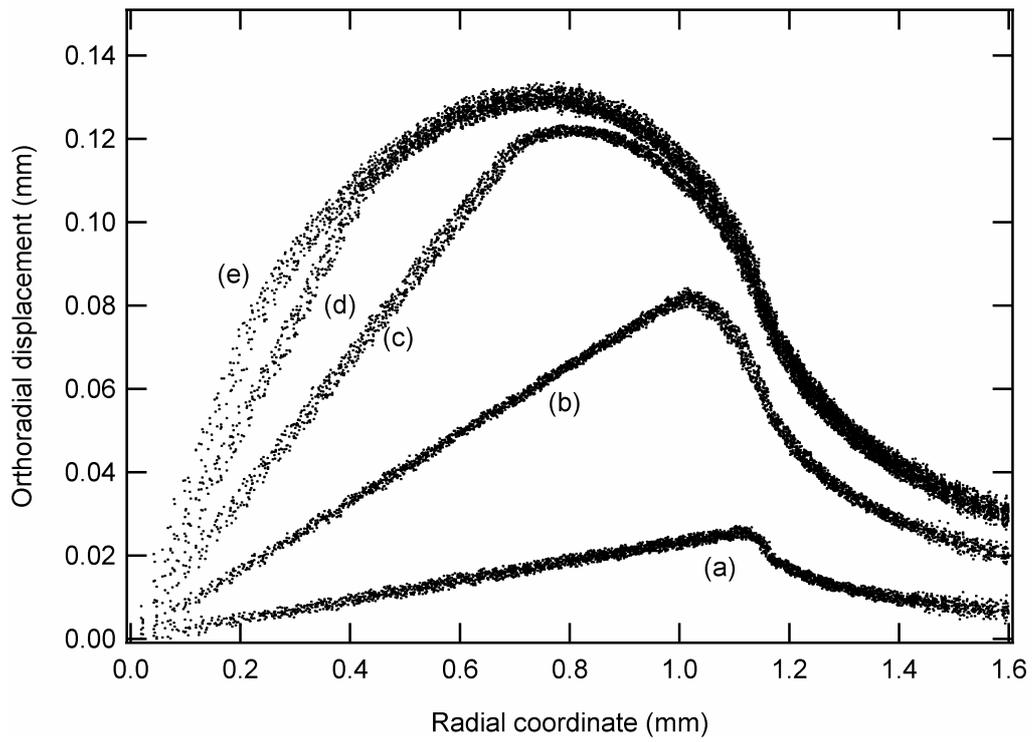

(I)

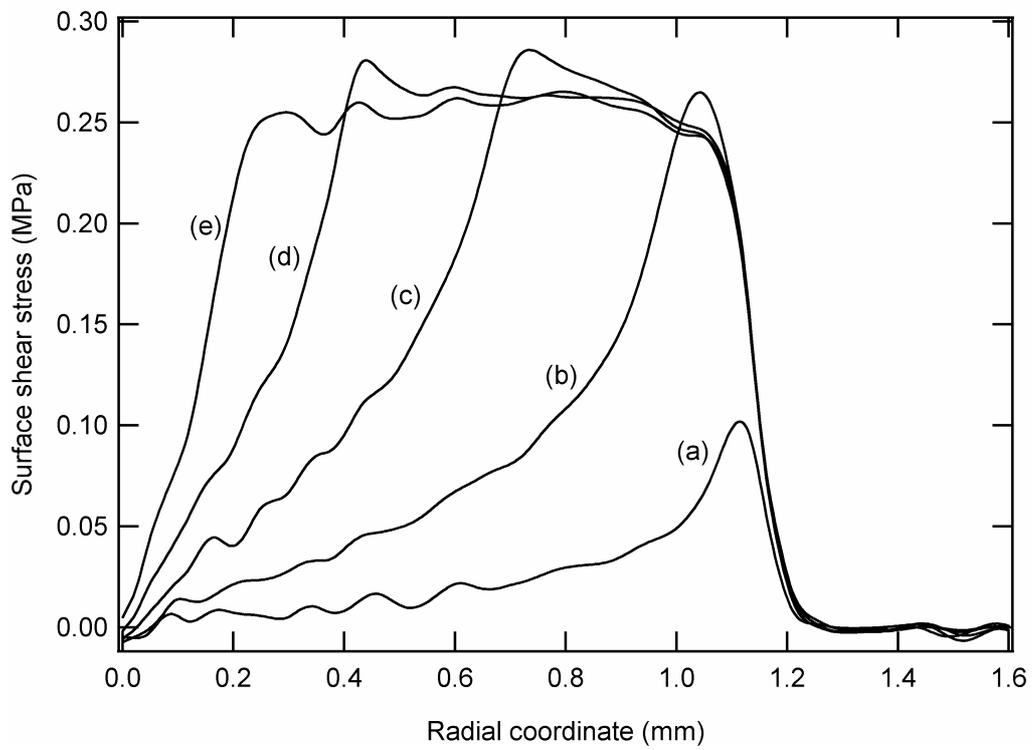

(II)